\documentstyle[prl,preprint,aps,epsf]{revtex}

\begin{document}
\bibliographystyle{plain}
\title{Ratio of absorption cross section for Dirac fermion to that
for scalar in the higher-dimensional black hole background}
\author{Eylee Jung\footnote{Email:eylee@kyungnam.ac.kr}, 
SungHoon Kim\footnote{Email:shoon@kyungnam.ac.kr} and
D. K. Park\footnote{Email:dkpark@hep.kyungnam.ac.kr 
}}
\address{Department of Physics, Kyungnam University,
Masan, 631-701, Korea.}
\date{\today}
\maketitle

\begin{abstract}
The ratio of the low-energy absorption cross section for Dirac fermion to 
that for minimally coupled scalar is computed when the spacetimes are
various types of the higher-dimensional Reissner-Nordstr\"{o}m black holes.
It is found that the low-energy absorption cross sections for the Dirac
fermion always goes to zero in the extremal limit regardless of the 
detailed geometry of the spacetime. The physical importance of our results
is discussed in the context of the brane-world scenarios and string theories. 
\end{abstract}
\newpage

Recent brane-world scenarios such as the large extra 
dimensions\cite{ark98-1,anto98} or warped extra dimensions\cite{rs99-1,rs99-2}
predict an emergence of the TeV-scale gravity. This crucial effect arising
due to the extra dimensions opens the possibility to make tiny black holes
in the future high-energy colliders\cite{gidd02-1,dimo01-1,eard02-1}. In this
reason much attention is paid recently to the absorption and emission problems
of the black holes in the presence of the extra dimensions.

If the extra dimensions exist, the absorption and emission problems should
be analyzed for the brane-localized and the bulk cases separately. It was
argued in Ref.\cite{argy98,banks99} that the Hawking radiation into the bulk
is dominant compared to that on the brane. This argument is mainly originated
from the fact that for the tiny black hole\footnote{We assume
$\ell_{fun} << r_0 << L$ where $\ell_{fun}$, $r_0$, and $L$ are bulk Planck
length, black hole radius, and size of the extra dimension.} the Hawking 
temperature is much larger than the mass of the light Kaluza-Klein modes. 
However, different argument was suggested by Emparan, Horowitz and 
Myers(EHM)\cite{emp00}. EHM argued that the radiation into the bulk by the 
light Kaluza-Klein modes is highly suppressed by the geometrical factor, which 
makes the {\it missing} energy much smaller than the {\it visible} one. 
EHM verified their argument explicitly by approximating the black hole as 
a black body with a critical radius $r_c$, which is slightly larger than
the horizon radius.
The detailed calculation on the absorption and emission problems for the 
$(4+n)$-dimensional Schwarzschild black hole was carried out 
analytically\cite{kanti02-1,kanti03-1,jung04} and numerically\cite{harris03-1},
which supports the main result of Ref.\cite{emp00}, {\it i.e.} the 
{\it visible} energy via the Hawking radiation is dominant. Especially, 
Ref.\cite{jung04} has shown via the analytic computation that the low-energy
absorption cross section for the brane-localized massive Dirac fermion over
that for the massive scalar is explicitly dependent on $n$, the number of 
the toroidally compactified extra dimensions, as $2^{(n-3) / (n+1)}$. Thus
the ratio factor $1/8$ is recovered when $n=0$, which was derived long ago
by Unruh\cite{unruh76}.

The absorption and emission problems of the higher-dimensional charged or
rotational black holes were also 
studied\cite{jung04-3,jung05-1,frol02-1,frol02-2,frol03-1,harris05-1}. 
For the charged
black holes the full absorption and emission spectra were computed
in Ref.\cite{jung05-1} by adopting the numerical technique used in 
Ref.\cite{sanc78,jung04-2}. It was shown that the increase of the inner 
horizon parameter, {\it say} $r_-$, in general enhances the absorptivity and
suppresses the emission rate compared to the case of the Schwarzschild
phase ($r_- = 0$) regardless of the brane-localized or bulk case. Also the 
bulk-to-brane energy emissivity was shown to decrease with increasing $r_-$.
The results of Ref.\cite{jung05-1} also support the main result of EHM if 
$n$ is not too large. However, for the large $n$ the emission into the bulk
can be dominant. For the rotating black holes it was shown in 
Ref.\cite{frol02-1,frol02-2,frol03-1,harris05-1} that the effect of 
the superradiance is 
very important for the experimental signature in the future collider. It was
argued that the rotating black holes, contrary to the Schwarzschild black hole,
radiate mainly into the bulk. More recently, the differences of the tiny
black holes arising due to the various extra dimensional scenarios are 
examined\cite{stoj04}. This may be helpful in the future black hole
experiment to determine a type of the extra dimensions.

In this letter we will compute the ratio of the low-energy absorption cross
section for Dirac fermion to that for the minimally coupled scalar when the
spacetimes are various types of the higher-dimensional 
Reissner-Nordstr\"{o}m(RN) black holes. We start with a brief review of 
Ref.\cite{das97} which examined the universality of the low-energy absorption
cross sections. Let us consider the general spherically symmetric metric
in $(p+2)$ spacetime dimensions of the form
\begin{equation}
\label{general1}
ds^2 = -f(r) dt^2 + g(r) \left[dr^2 + r^2 d\Omega_p^2\right]
\end{equation}
where $d\Omega_p^2$ is a spherically symmetric angular part in the metric.
We also assume 
\begin{equation}
\label{asympflat}
\lim_{r \rightarrow \infty} f(r) = 
\lim_{r \rightarrow \infty} g(r) = 1
\end{equation}
for imposing the asymptotically flat. Then it is well-known that the low-energy
absorption cross section for the minimally coupled bulk scalar is equal to
the horizon area
\begin{equation}
\label{blscalar}
\sigma_S = \Omega_p R_H^p \equiv {\cal A}_H
\end{equation}
where $\Omega_p = 2 \pi^{(p+1)/2} / \Gamma (p+1/2)$, 
$R_H = r_H \sqrt{g(r_H)}$, and $r_H$ is a horizon radius usually 
determined by the largest solution of $f(r)=0$.  
Furthermore Ref.\cite{das97} has shown that the low-energy
absorption cross section for the bulk fermion is given by the area measured 
in the flat spatial metric conformally related to the true metric in the form
\begin{equation}
\label{blfermion}
\sigma_F = 2 \Omega_p r_H^p = 2 \left(g(r_H)\right)^{-\frac{p}{2}}
                                  {\cal A}_H
\end{equation}
where the factor $2$ comes from the number of spinors. Thus for the bulk fields
we can show the ratio factor $\sigma_F / \sigma_S$ to be
\begin{equation}
\label{ratio1}
\gamma^{BL} = \frac{\sigma_F}{\sigma_S} = 2 \left(g(r_H)\right)^{-\frac{p}{2}}.
\end{equation}
If we assume that the three brane has negligible self-gravity of its own, 
the induced metric on the brane can be written as 
\begin{equation}
\label{general2}
ds^2 = -f(r) dt^2 + g(r) \left[dr^2 + r^2 d\Omega_2^2\right].
\end{equation}
Following the same procedure, it is easy to show that the low-energy absorption
cross section for the brane-localized scalar is $\sigma_S = 4 \pi R_H^2$ and 
the ratio factor is 
\begin{equation}
\label{ratio2}
\gamma^{BR} = \frac{\sigma_F}{\sigma_S} = 2 \left(g(r_H)\right)^{-1}.
\end{equation}

Now, we consider the $(4+n)$-dimensional charged black hole whose 
metric is\cite{myers86} 
\begin{equation}
\label{space1}
ds_{RN}^2 = -\left[1 - \left(\frac{\tilde{r}_+}{\tilde{r}}\right)^{n+1} \right]
        \left[1 - \left(\frac{\tilde{r}_-}{\tilde{r}}\right)^{n+1} \right] dt^2
      + \frac{d \tilde{r}^2}
             {\left[1 - \left(\frac{\tilde{r}_+}{\tilde{r}}\right)^{n+1} \right]        \left[1 - \left(\frac{\tilde{r}_-}{\tilde{r}}\right)^{n+1} \right]}
      + \tilde{r}^2 d \Omega_{n+2}^2
\end{equation}
where $\tilde{r}_+$ and $\tilde{r}_-$ correspond to the external and internal 
horizons, respectively.
It is convenient to introduce a new radial coordinate $\hat{r}$ in the form
\begin{equation}
\label{change1}
\tilde{r}^2 = a(\hat{r}) \hat{r}^2
\end{equation}
where 
\begin{equation}
\label{def1}
a(\hat{r}) = \left[ 1 + \left( \frac{\hat{r}_1}{\hat{r}} \right)^{n+1}
                                      \right]^{\frac{2}{n+1}}
\end{equation}
and $\hat{r}_1 \equiv \tilde{r}_-$. Then the metric (\ref{space1}) reduces to 
the following more tractable form
\begin{equation}
\label{space2}
ds_{RN}^2 = -h(\hat{r}) a^{-n-1}(\hat{r}) dt^2 + a(\hat{r})
\left[h^{-1}(\hat{r}) d\hat{r}^2 + \hat{r}^2 d\Omega_{n+2}^2\right]
\end{equation}
where 
\begin{equation}
\label{def2}
h(\hat{r}) = 1 - \left( \frac{\hat{r}_0}{\hat{r}}\right)^{n+1}
\end{equation}
and $\hat{r}_0^{n+1} = \tilde{r}_+^{n+1} - \tilde{r}_-^{n+1}$. It is easy
to show that the horizon radius $\tilde{r} = \tilde{r}_+$ in 
$\tilde{r}$-coordinate corresponds to $\hat{r} = \hat{r}_0$ in 
$\hat{r}$-coordinate.

The metric (\ref{space2}) can be written in the form
\begin{equation}
\label{space3}
ds_{RN}^2 = -f_{RN}(r) dt^2 + 
g_{RN}(r) \left[dr^2 + r^2 d\Omega_{n+2}^2 \right]
\end{equation}
if $f_{RN}(r)$, $g_{RN}(r)$ and $r$ satisfy
\begin{eqnarray}
\label{rela1}
& &f_{RN}(r) = h\left(\hat{r}(r)\right) a^{-n-1}\left(\hat{r}(r)\right)
                                                          \\  \nonumber
& &g_{RN}(r) = \frac{\hat{r}^2}{r^2} a\left(\hat{r}(r)\right)
                                                          \\  \nonumber
& &\sqrt{g_{RN}(r)} dr = \sqrt{\frac{a(\hat{r})}{h(\hat{r})}} d\hat{r}.
\end{eqnarray}
Solving Eq.(\ref{rela1}), one can show straightforwardly that the $r$-dependence
of $\hat{r}$ is 
\begin{equation}
\label{coord1}
\hat{r} = \hat{r}_0 \frac{{\cal C}}{r}
\left[\frac{2}{\left(\frac{r}{{\cal C}}\right)^{n+1} + 1} \right]^{-\frac{2}
                                                                   {n+1}}.
\end{equation}
where ${\cal C}$ is an integration constant. Eq.(\ref{coord1}) implies
\begin{equation}
\label{coord2}
h(\hat{r}) = \left(\frac{\left(\frac{r}{{\cal C}}\right)^{n+1} - 1}
                        {\left(\frac{r}{{\cal C}}\right)^{n+1} + 1}
                                           \right)^2
\hspace{1.0cm}
a(\hat{r}) = \left[1 + \frac{4 \left(\frac{\hat{r}_1}{\hat{r}_0}\right)^{n+1}
                             \left(\frac{r}{{\cal C}}\right)^{n+1} }
                            {\left[\left(\frac{r}{{\cal C}}\right)^{n+1} + 1
                                        \right]^2}
                                                     \right]^{\frac{2}{n+1}}.
\end{equation}
Combining (\ref{rela1}), (\ref{coord1}) and (\ref{coord2}), one can readily
derive 
\begin{eqnarray}
\label{coord3}
f_{RN}(r)&=&
\left( \frac{ \left[ \left( \frac{r}{{\cal C}} \right)^{n+1} + 1 \right]
              \left[ \left( \frac{r}{{\cal C}} \right)^{n+1} - 1 \right]
            }
            { \left[ \left( \frac{r}{{\cal C}} \right)^{n+1} + 1 \right]^2
             + 4 \left(\frac{\hat{r}_1}{\hat{r}_0}\right)^{n+1}
                             \left(\frac{r}{{\cal C}}\right)^{n+1}
            } 
                              \right)^2
                                                     \\  \nonumber
g_{RN}(r)&=& \left[ \left(\frac{{\cal C} \hat{r}_0}{r^2} \right)^{n+1}
                \left[\frac{\left( \frac{r}{{\cal C}} \right)^{n+1} + 1}{2}
                                                  \right]^2 + 
                    \left(\frac{\hat{r}_1}{r}\right)^{n+1} 
                                        \right]^{\frac{2}{n+1}}.
\end{eqnarray}
The integration constant ${\cal C}$ should be 
${\cal C} = 2^{-2 / (n+1)} \hat{r}_0$ due to Eq.(\ref{asympflat}). Thus 
$\hat{r} = \hat{r}_0$ in Eq.(\ref{coord1}) implies
\begin{equation}
\label{horizon2}
r_H = {\cal C} = 2^{-\frac{2}{n+1}} \hat{r}_0.
\end{equation}
Inserting Eq.(\ref{horizon2}) into (\ref{coord3}) yields
\begin{equation}
\label{horizon3}
g_{RN}(r_H) = 2^{\frac{4}{n+1}}
             \frac{\left[\hat{r}_0^{n+1} + \hat{r}_1^{n+1}\right]^{
                                                             \frac{2}{n+1}}
                  }{\hat{r}_0^2}
            = 2^{\frac{4}{n+1}}
              \frac{\tilde{r}_+^2}
                   {\left[\tilde{r}_+^{n+1} - \tilde{r}_-^{n+1} \right]^{
                                                              \frac{2}{n+1}}
                   },
\end{equation}
which makes $R_H = r_H \sqrt{g(r_H)} = \tilde{r}_+$. Thus the low-energy
absorption cross section for the bulk scalar should be 
$\sigma_S^{BL} = \Omega_{n+2} \tilde{r}^{n+2}$ and the ratio factor
$\gamma^{BL}$ given in Eq.(\ref{ratio1}) becomes
\begin{equation}
\label{ratio3}
\gamma^{BL} \equiv \frac{\sigma_F^{BL}}{\sigma_S^{BL}}
            = 2^{-\frac{n+3}{n+1}}
             \left[1 - \left(\frac{\tilde{r}_-}{\tilde{r}_+}\right)^{n+1}
                           \right]^{\frac{n+2}{n+1}}.
\end{equation}
Thus increasing $\tilde{r}_-$ decreases $\gamma^{BL}$ and eventually it goes
to zero at the extremal limit ($\tilde{r}_- = \tilde{r}_+$)\footnote{Since the 
black hole charge is 
$Q = \pm (\tilde{r}_+ \tilde{r}_-)^{(n+1)/2} \sqrt{(n+1)(n+2)/8 \pi}$, 
increase of $\tilde{r}_-$ means increase of the black hole charge. if it
is possible to fix the black hole entropy during increase of the black hole
charge, one can increase $\tilde{r}_-$ with fixed $\tilde{r}_+$}. 
In the Schwarzschild limit ($\tilde{r}_- = 0$)
$\gamma^{BL}$ becomes $2^{-(n+3) / (n+1)}$ which recovers $1/8$ at $n=0$.
Contrary to $\tilde{r}_-$, the increase of $n$ with fixed $\tilde{r}_-$ 
increases $\gamma^{BL}$ and eventually is saturated to $0.5$ when $n$ is 
infinity. For small $\tilde{r}_-$ the $n$-dependence of $\gamma^{BL}$ reaches
the saturated value more rapidly compared to the large $\tilde{r}_-$ case.

For the brane-localized scalar the low-energy absorption cross section is 
$\sigma_S^{BR} = 4 \pi \tilde{r}_+^2$ and the ratio factor reduces to 
\begin{equation}
\label{ratio4}
\gamma^{BR} \equiv \frac{\sigma_F^{BR}}{\sigma_S^{BR}}
            = 2^{\frac{n-3}{n+1}} 
                \left[1 - \left(\frac{\tilde{r}_-}{\tilde{r}_+}\right)^{n+1}
                           \right]^{\frac{2}{n+1}}.
\end{equation}
As in the case of the bulk fields $\gamma^{BR}$ approaches to zero at the 
extremal limit\footnote{Following Ref.\cite{jung04}, we can derive 
Eq.(\ref{ratio4}) by direct matching procedure. Since it is tedious and lengthy
procedure, it is not presented in this letter.}. 
In the Schwarzschild limit Eq.(\ref{ratio4}) reproduces 
$\gamma^{BR} = 2^{(n-3)/(n+1)}$, which was derived via the direct matching 
procedure in Ref.\cite{jung04}. The $n$-dependence of $\gamma^{BR}$ is similar
to that of $\gamma^{BL}$, but the saturated value is changed into $2$. The 
increas of $\gamma^{BR}$ with increasing $n$ is consistent with Fig. 3 of 
Ref.\cite{harris03-1}.

Next we consider a five-dimensional black hole with three different 
$U(1)$ charges\cite{cve96,horo96}
\begin{equation}
\label{malda1}
ds_{MS}^2 = -h(\tilde{r}) a^{-2/3}(\tilde{r}) dt^2 + a^{1/3}(\tilde{r})
           \left[h^{-1}(\tilde{r}) d \tilde{r}^2 + 
           \tilde{r}^2 d \Omega_3^2 \right]
\end{equation}
where
\begin{eqnarray}
\label{defmal1}
& &h(\tilde{r}) = 1 - \frac{\tilde{r}_0^2}{\tilde{r}^2}
\hspace{2.0cm}
a(\tilde{r}) = a_1(\tilde{r}) a_5(\tilde{r}) a_K(\tilde{r})
                                                 \\  \nonumber
& &
\hspace{2.0cm}
a_i(\tilde{r}) = 1 + \frac{\tilde{r}_i^2}{\tilde{r}^2}
\hspace{1.0cm}
(i = 1, 5, K).
\end{eqnarray}
The $U(1)$ charges $Q_i = (\tilde{r}_0^2 / 2) \sinh 2 \xi_i$ where
$\xi_i = \sinh^{-1} (\tilde{r}_i / \tilde{r}_0)$ are originated from
1D-branes, 5D-branes and Kaluza-Klein charges respectively. The spacetime
(\ref{malda1}) is extensively used by Maldacena and Strominger in 
Ref.\cite{mal96-1} for the examination of the black hole-D-brane 
correspondence in the dilute gas region $\tilde{r}_0 << \tilde{r}_K <<
\tilde{r}_1, \tilde{r}_5$ and by Hawking and Taylor-Robinson\cite{hawk97} in
the slightly different region $\tilde{r}_0 << \tilde{r}_1, \tilde{r}_5,
\tilde{r}_K$. Note that the spacetime (\ref{malda1}) exactly 
coincides with an usual five-dimensional RN black hole when 
$\tilde{r}_1 = \tilde{r}_5 = \tilde{r}_K$.

The metric (\ref{malda1}) can be re-written in the form
\begin{equation}
\label{metric10}
ds_{MS}^2 = -f_D(r) dt^2 + g_D(r) (dr^2 + r^2 d\Omega_3^2)
\end{equation}
if $f_D(r)$, $g_D(r)$, and $r$ satisfy
\begin{eqnarray}
\label{rela2}
f_D(r)&=&h(\tilde{r}) a^{-2/3}(\tilde{r})
                                         \\   \nonumber
g_D(r)&=&\frac{\tilde{r}^2}{r^2} a^{1/3}(\tilde{r})
                                         \\   \nonumber
\sqrt{g_D(r)} dr&=&\frac{a^{1/6}(\tilde{r})}{\sqrt{h(\tilde{r})}}
d \tilde{r}.
\end{eqnarray}
Solving Eq.(\ref{rela2}) yields the following $r$-dependence of 
$\tilde{r}$;
\begin{equation}
\label{depen1}
\tilde{r} = \tilde{r}_0 
           \frac{{\cal C}}{2 r}
           \left[ \frac{r^2}{{\cal C}^2} + 1 \right]
\end{equation}
where ${\cal C}$ is an integration constant. Eq.(\ref{depen1}) implies
\begin{eqnarray}
\label{rela3}
h(\tilde{r})&=&\left( \frac{\frac{r^2}{{\cal C}^2} - 1}
                           {\frac{r^2}{{\cal C}^2} + 1} \right)^2
                                                    \\   \nonumber
a(\tilde{r})&=&\left[ 1 + \frac{4 
                                \left(\frac{\tilde{r}_1}{\tilde{r}_0} \right)^2
                                \left(\frac{r}{{\cal C}}\right)^2
                               }
                              {\left(\frac{r^2}{{\cal C}^2} + 1 \right)^2
                               }
                                  \right]
                \left[ 1 + \frac{4 
                                \left(\frac{\tilde{r}_5}{\tilde{r}_0} \right)^2
                                \left(\frac{r}{{\cal C}}\right)^2
                               }
                              {\left(\frac{r^2}{{\cal C}^2} + 1 \right)^2
                               }
                                  \right]
                \left[ 1 + \frac{4 
                                \left(\frac{\tilde{r}_K}{\tilde{r}_0} \right)^2
                                \left(\frac{r}{{\cal C}}\right)^2
                               }
                              {\left(\frac{r^2}{{\cal C}^2} + 1 \right)^2
                               }
                                  \right].
\end{eqnarray}
Combining (\ref{rela2}), (\ref{depen1}) and (\ref{rela3}), one can derive
\begin{eqnarray}
\label{rela4}
f_D(r)&=&\left( \frac{r^2}{{\cal C}^2} + 1 \right)^2
                       \left( \frac{r^2}{{\cal C}^2} - 1 \right)^2
                      \left\{ \left( \frac{r^2}{{\cal C}^2} + 1 \right)^2
                               + 4 
                                \left(\frac{\tilde{r}_1}{\tilde{r}_0} \right)^2
                                \left(\frac{r}{{\cal C}}\right)^2
                        \right\}^{-2/3}
                                               \\  \nonumber
& & \times
                        \left\{ \left( \frac{r^2}{{\cal C}^2} + 1 \right)^2
                               + 4 
                                \left(\frac{\tilde{r}_5}{\tilde{r}_0} \right)^2
                                \left(\frac{r}{{\cal C}}\right)^2
                        \right\}^{-2/3}
                        \left\{ \left( \frac{r^2}{{\cal C}^2} + 1 \right)^2
                               + 4 
                                \left(\frac{\tilde{r}_K}{\tilde{r}_0} \right)^2
                                \left(\frac{r}{{\cal C}}\right)^2
                        \right\}^{-2/3}
                                               \\   \nonumber
g_D(r)&=&\left[\frac{{\cal C}^2 \tilde{r}_0^2}
                    {4 r^4}
                    \left( \frac{r^2}{{\cal C}^2} + 1 \right)^2
                   + \left(\frac{\tilde{r}_1}{r}\right)^2 \right]^{1/3}
         \left[\frac{{\cal C}^2 \tilde{r}_0^2}
                    {4 r^4}
                    \left( \frac{r^2}{{\cal C}^2} + 1 \right)^2
                   + \left(\frac{\tilde{r}_5}{r}\right)^2 \right]^{1/3}
                                                \\   \nonumber
& &\hspace{6.0cm} \times
         \left[\frac{{\cal C}^2 \tilde{r}_0^2}
                    {4 r^4}
                    \left( \frac{r^2}{{\cal C}^2} + 1 \right)^2
                   + \left(\frac{\tilde{r}_K}{r}\right)^2 \right]^{1/3}.
\end{eqnarray}
From a condition of the asymptotic flat the integration constant
${\cal C}$ is fixed as ${\cal C} = \tilde{r}_0 / 2$. Since the location
of the horizon is $\tilde{r} = \tilde{r}_0$ in $\tilde{r}$-coordinate,
Eq.(\ref{depen1}) indicates that the horizon radius in $r$-coordinate is 
\begin{equation}
\label{horizon4}
r \equiv \frac{\tilde{r}_0}{2} = r_H.
\end{equation}
Inserting Eq.(\ref{horizon4}) into (\ref{rela4}), one can show easily
\begin{eqnarray}
\label{horizon5}
g_D(r_H)&=&4 \left[1 + \left( \frac{\tilde{r}_1}{\tilde{r}_0} \right)^2
                                                    \right]^{1/3}
             \left[1 + \left( \frac{\tilde{r}_5}{\tilde{r}_0} \right)^2
                                                    \right]^{1/3}
             \left[1 + \left( \frac{\tilde{r}_K}{\tilde{r}_0} \right)^2
                                                    \right]^{1/3}
                                                        \\  \nonumber
R_H&\equiv& r_H \sqrt{g_D(r_H)} = 
\left(\tilde{r}_0^2 + \tilde{r}_1^2\right)^{1/6}
\left(\tilde{r}_0^2 + \tilde{r}_5^2\right)^{1/6}
\left(\tilde{r}_0^2 + \tilde{r}_K^2\right)^{1/6}.
\end{eqnarray}
Thus Eq.(\ref{blscalar}) with $p=3$ yields the low-energy absorption cross 
section for the massless bulk scalar
\begin{equation}
\label{absbls1}
\sigma_S^{BL}=2\pi^2 R_H^3=2\pi^2 (\tilde{r}_0^2 +\tilde{r}_1^2)^{\frac{1}{2}}
              (\tilde{r}_0^2 +\tilde{r}_5^2)^{\frac{1}{2}}
              (\tilde{r}_0^2 +\tilde{r}_K^2)^{\frac{1}{2}}.
\end{equation}
which exactly equals to the horizon area.

In the dliute gas region ($\tilde{r}_0 \ll \tilde{r}_K \ll \tilde{r}_1,
\tilde{r}_5 $) Eq.(\ref{absbls1}) can be written in the form
\begin{equation}
\label{absbls2}
\sigma_S^{BL} \sim 2 \pi^2 \tilde{r}_0 \tilde{r}_1 \tilde{r}_5
             \sqrt{1 + \left( \frac{\tilde{r}_K}{\tilde{r}_0} \right)^2}
             \sim 2 \pi^2 \tilde{r}_1^2 \tilde{r}_5^2 \frac{\pi \omega}{2}
                 \frac{e^{\frac{\omega}{T_H}}-1}{(e^{\frac{\omega}{2T_L}}-1)
                    (e^{\frac{\omega}{2T_R}}-1)}
\end{equation}
where $T_L$ and $T_R$, the temperatures of left and right moving
oscillations, are related to the Hawking temperature $T_H$ by
\footnote{In the dilute gas region 
$T_H \sim \tilde{r}_0 /2 \pi \tilde{r}_1 \tilde{r}_5 \cosh \zeta$, 
$T_L \sim \tilde{r}_0 e^{\zeta} /2 \pi \tilde{r}_1 \tilde{r}_5$ and 
$T_R \sim \tilde{r}_0 e^{-\zeta} /2 \pi \tilde{r}_1 \tilde{r}_5$
where $\zeta = \sinh^{-1}(\tilde{r}_K / \tilde{r}_0)$.}
\begin{equation}
\label{hawkingth}
\frac{1}{T_R} + \frac{1}{T_L} = \frac{2}{T_H}.
\end{equation}
Eq.(\ref{ratio1}) with $p=3$ gives a ratio factor
\begin{equation}
\label{ratiobl1}
\gamma_D^{BL} \equiv \frac{\sigma_F^{BL}}{\sigma_S^{BL}} 
= \frac{1}{4} \left[1 + \left( \frac{\tilde{r}_1}{\tilde{r}_0} \right)^2
                                                    \right]^{-\frac{1}{2}}
             \left[1 + \left( \frac{\tilde{r}_5}{\tilde{r}_0} \right)^2
                                                    \right]^{-\frac{1}{2}}
             \left[1 + \left( \frac{\tilde{r}_K}{\tilde{r}_0} \right)^2
                                                    \right]^{-\frac{1}{2}}.
\end{equation}
It is interesting to note that Eq.(\ref{ratiobl1}) indicates
$\sigma_F^{BL}$ is independent of $\tilde{r}_1,\tilde{r}_5$ and $\tilde{r}_K$.
If one angle variable in Eq.(\ref{malda1}) is regarded as a toroidally 
compactified extra dimension, one can show using Eq.(\ref{ratio2}) that
the low-energy absorption cross section for the brane-localized scalar is
\begin{equation}
\label{absbrs1}
\sigma_S^{BR} = 4 \pi^2 R_H^2 
= 4 \pi (\tilde{r}_0^2 +\tilde{r}_1^2)^{\frac{1}{3}}
              (\tilde{r}_0^2 +\tilde{r}_5^2)^{\frac{1}{3}}
              (\tilde{r}_0^2 +\tilde{r}_K^2)^{\frac{1}{3}}
\end{equation}
and the ratio factor is
\begin{equation}
\label{ratiobr1}
\gamma_D^{BR} \equiv \frac{\sigma_F^{BR}}{\sigma_S^{BR}} 
= 2 g^{-1}(r_H)
= \frac{1}{2} \left[1 + \left( \frac{\tilde{r}_1}{\tilde{r}_0} \right)^2
                                                    \right]^{-\frac{1}{3}}
             \left[1 + \left( \frac{\tilde{r}_5}{\tilde{r}_0} \right)^2
                                                    \right]^{-\frac{1}{3}}
             \left[1 + \left( \frac{\tilde{r}_K}{\tilde{r}_0} \right)^2
                                                    \right]^{-\frac{1}{3}}.
\end{equation}
Eq.(\ref{ratiobr1}) also indicates that $\sigma_F^{BR}$ is independent of 
$\tilde{r}_1,\tilde{r}_5$ and $\tilde{r}_K$.

In this letter we calculated the ratio of the low-energy cross section for 
Dirac fermion to that for scalar when the spacetimes are various types of the
higher-dimensional RN black holes. It was found that the low-energy
cross section for Dirac fermion always goes to zero in the extremal limit.
One may confirm our results (\ref{ratio3}) and (\ref{ratio4}) by 
computing the absorption cross section in the full energy range numerically
by adopting the numerical technique used in Ref.\cite{jung04}. Also it is 
interesting to check whether or not the D-brane approach\cite{das96} derives
Eq.(\ref{ratiobl1}) and (\ref{ratiobr1}). Among them it may be of greatest
interest to extend our calculation to the non-spherically symmetric spacetimes
such as the Kerr-Newman black holes.

\vspace{1cm}

{\bf Acknowledgement}:  
This work was supported by the Kyungnam University
Research Fund, 2004.


\begin{thebibliography}{99}
\bibitem{ark98-1} N. Arkani-Hamed, S. Dimopoulos and G. Dvali,
{\it The Hierarchy Problem and New Dimensions at a Millimeter}, 
Phys. Lett. {\bf B429} (1998) 263 [hep-ph/9803315].
\bibitem{anto98} L. Antoniadis, N. Arkani-Hamed, S. Dimopoulos and G. Dvali,
{\it New Dimensions at a Millimeter to a Fermi and Superstrings at a 
TeV}, Phys. Lett. {\bf B436} (1998) 257 [hep-ph/9804398]. 
\bibitem{rs99-1} L. Randall and R. Sundrum, {\it A Large Mass Hierarchy from a 
Small Extra Dimension}, 
Phys. Rev. Lett. {\bf 83} (1999) 3370 [hep-ph/9905221].
\bibitem{rs99-2} L. Randall and R. Sundrum, {\it An Alternative to 
Compactification}, Phys. Rev. Lett. {\bf 83} (1999) 4690 [hep-th/9906064].
\bibitem{gidd02-1} S. B. Giddings and T. Thomas, {\it High energy colliders 
as black hole factories: The end of short distance physics}, Phys. Rev. 
{\bf D65} (2002) 056010 [hep-ph/0106219].
\bibitem{dimo01-1} S. Dimopoulos and G. Landsberg, {\it Black Holes at the 
Large Hadron Collider}, Phys. Rev. Lett. {\bf 87} (2001) 161602 
[hep-ph/0106295].
\bibitem{eard02-1} D. M. Eardley and S. B. Giddings, {\it Classical black hole 
production in high-energy collisions}, Phys. Rev. {\bf D66} (2002)
044011 [gr-qc/0201034].
\bibitem{argy98} P. Argyres, S. Dimopoulos and J. March-Russell, 
{\it Black Holes and Sub-millimeter Dimensions}, Phys. Lett. 
{\bf B441} (1998) 96 [hep-th/9808138].
\bibitem{banks99} T. Banks and W. Fischler, {\it A Model for High Energy 
Scattering
in Quantum Gravity} [hep-th/9906038].
\bibitem{emp00} R. Emparan, G. T. Horowitz and R. C. Myers, {\it Black Holes
radiate mainly on the Brane}, Phys. Rev. Lett. {\bf 85} (2000) 499
[hep-th/0003118].
\bibitem{kanti02-1} P. Kanti and J. March-Russel, {\it Calculable corrections 
to brane black hole decay: The scalar case}, Phys. Rev. {\bf D66} (2002)
024023 [hep-ph/0203223] 
\bibitem{kanti03-1} P. Kanti and J. March-Russel, {\it Calculable corrections 
to brane black hole decay: II. Greybody factors for spin $1/2$ and $1$}, 
Phys. Rev. {\bf D67} (2003) 104019 [hep-ph/0212199].
\bibitem{jung04} E. Jung, S. H. Kim, and D. K. Park, {\it Low-energy absorption
cross section for massive scalar and Dirac fermion by $(4+n)$-dimensional 
Schwarzschild black hole}, JHEP {\bf 0409} (2004) 005 [hep-th/0406117]. 
\bibitem{harris03-1} C. M. Harris and P. Kanti, {\it Hawking Radiation from a
$(4+n)$-dimensional Black Hole: Exact Results for the Schwarzschild Phase},
JHEP {\bf 0310} (2003) 014 [hep-ph/0309054].
\bibitem{unruh76} W. G. Unruh, {\it Absorption cross section of small 
black holes},
Phys. Rev. {\bf D14} (1976) 3251.
\bibitem{jung04-3} E. Jung, S. H. Kim, and D. K. Park, {\it Proof of 
universality for the absorption of massive scalar by the higher-dimensional
Reissner-Nordstr\"{o}m black holes}, Phys. Lett. {\bf B602} (2004)
105 [hep-th/0409145].
\bibitem{jung05-1} E. Jung and D. K. Park, {\it Absorption and Emission 
Spectra of an higher-dimensional Reissner-Nordstr\"{o}m black hole}
[hep-th/0502002].
\bibitem{frol02-1} V. Frolov and D. Stojkovi\'{c}, {\it Black hole radiation
in the brane world and the recoil effect}, Phys. Rev. {\bf D66} (2002)
084002 [hep-th/0206046].
\bibitem{frol02-2} V. Frolov and D. Stojkovi\'{c}, {\it Black Hole as a Point
Radiator and Recoil Effect on the Brane World}, Phys. Rev. Lett. 
{\bf 89} (2002) 151302 [hep-th/0208102].
\bibitem{frol03-1} V. Frolov and D. Stojkovi\'{c}, {\it Quantum radiation from 
a $5$-dimensional black hole}, Phys. Rev. {\bf D67} (2003) 084004
[gr-qc/0211055].
\bibitem{harris05-1} C. M. Harris and P. Kanti, {\it Hawking Radiation from a 
$(4+n)$-Dimensional Rotating Black Hole} [hep-th/0503010].
\bibitem{sanc78} N. Sanchez, {\it Absorption and emission spectra of a
Schwarzschild black hole}, Phys. Rev. {\bf D18} (1978) 1030.
\bibitem{jung04-2} E. Jung and D. K. Park, {\it Effect of Scalar Mass in the 
Absorption and Emission Spectra of Schwarzschild Black Hole}, Class. Quant.
Grav. {\bf 21} (2004) 3717 [hep-th/0403251].
\bibitem{stoj04} D. Stojkovic, {\it Distinguishing between the small ADD and
RS black holes in accelerators}, Phys. Rev. Lett. {\bf 94} (2005)
011603 [hep-ph/0409124].
\bibitem{das97} S. R. Das, G. Gibbons, and S. D. Mathur, 
{\it Universality of Low Energy Absorption Cross Sections for Black Holes},
Phys. Rev. Lett. {\bf 78} (1997) 417 [hep-th/9609052].
\bibitem{myers86} R. C. Myers and M. J. Perry, {\it Black Holes in Higher
Dimensional Space-Times}, Ann. Phys. {\bf 172} (1986) 304.
\bibitem{cve96} M. Cvetic and D. Youm, {\it General Rotating Five Dimensional 
Black Holes of Toroidally Compactified Heterotic String}, Nucl. Phys.
{\bf B476} (1996) 118 [hep-th/9603100].
\bibitem{horo96} G. Horowitz, J. Maldacena, and A. Strominger, 
{\it Nonextremal Black Hole Microstates and U-duality}, Phys. Lett. 
{\bf B383} (1996) 151 [hep-th/9603109].
\bibitem{mal96-1} J. Maldacena and A. Strominger, {\it Black Hole Greybody
Factors and D-Brane Spectroscopy}, Phys. Rev. {\bf D55} (1997) 861
[hep-th/9609026].
\bibitem{hawk97} S. W. Hawking and M. M. Taylor-Robinson, {\it Evolution of 
near-extremal black holes}, Phys. Rev. {\bf D55} (1997) 7680
[hep-th/9702045].
\bibitem{das96} S. R. Das and S. D. Mathur, {\it Comparing decay rates for 
black holes and D-branes}, Nucl. Phys. {\bf B478} (1996) 561
[hep-th/9606185].



\end{thebibliography}
\end{document}